\documentclass{ptapap}
\usepackage{amsmath}
\usepackage{amssymb}

\author{Ishika Palit}[CFT-PAS]
\author{Agnieszka Janiuk}[CFT-PAS]
\author{Petra Sukova}[ASU-CAS]
\affil[CFT-PAS]{Center for Theoretical Physics of Polish Academy of Sciences, Al. Lotnikow 32/46, 02-668 Warsaw, Poland}
\affil[ASU-CAS]{Astronomical Institute, Czech Academy of Sciences, Bocni II 1401, 141 00 Prague, Czech Republic}

\title{Mass and spin constraint on black holes in long GRBs}

\begin{document}
\maketitle

\begin{abstract}
We compute the evolution of a quasi-spherical, slowly rotating accretion flow around a black hole, whose mass and spin evolve adequately to the mass-energy transfer through the horizon. Our model is relevant for the central engine driving a long gamma ray burst, that originates from the collapse of a massive star. Our results show how much mass and spin a newly formed black hole should possess during collapsar to launch long GRB. 

\end{abstract}

\section{Introduction}

Gamma ray bursts are highly energetic and brightest explosions that
have been observed in EM spectrum. These can last from few seconds to few hours. The progenitors of long GRBs are believed to be massive stars exploding due to the collapse of their cores. Matter from the star around the core falls down towards the center forming a gaseous envelope and (for rapidly rotating stars) swirls into a high density accretion disk. The computations of a GRB engine in a dynamically evolving spacetime metric are important specifically due to the transient nature of the event, in which a huge amount of mass is accreted and changes the fundamental black hole parameters, its mass and spin, during the process. 

\section{Model description}
We start with a newly formed black hole whose mass and spin are going to evolve depending on the rotation of the collapsing cloud.
We set a critical angular momentum of the cloud at certain circularisation radii and further study the growth of black hole in sub critical, critical and super critical regime. The stellar structure is described using a slowly rotating, quasi-spherical flow, with the relativistic solution for the Bondi-Michel radial dependence of density, and specific angular momentum concentrated
at the equator \citep{2008ApJ...687..433J} .

\subsection{Evolution of mass and spin:}

In simulation, the evolution of the black hole mass M and spin J is computed according to the equations in \cite{2004ApJ...602..312G}. The changing black hole spin and mass are subsequently affecting the spacetime metric. The mass growth is updated in every time step according to $ \Delta M = \frac{M ^{curr}_{BH}} {M ^{0}_{BH}}  - 1$ where $M ^{0}_{BH}$ is the initial mass of the black hole, and the current mass is given by integration of the rest-mass flux over the horizon at every time-step: 
\begin{equation}
 M ^{curr}_{BH} = M ^{i}_{BH} = M ^{i-1}_{BH} + \int_{r = r_{in}} dM_{in}2 \pi d\theta \sqrt{-g} \Delta t   
\end{equation}  
where $dM_{in} = - \rho\frac{u^{r}}{u^{t}}$ .
The change of the spin parameter of the black hole is computed as:
\begin{equation}
 a^{i} = a^{i-1} + (\frac{\dot{J}}{M ^{curr}_{BH}} - \frac{a^{i-1}}{M ^{curr}_{BH}} \dot{E})\Delta t 
\end{equation}
Our models are scaled with a parameter defining the ratio between our collapsar’s
angular momentum and the critical angular momentum value at the circular orbit:
\begin{equation}
l_{\rm spec} = S |\frac{u_{\phi}}{u_{t}}| \sin^{2}\theta
\end{equation} 
with $u^{\phi}$ defined as $u^{\phi} = g^{t\phi} (-\epsilon) +g^{\phi\phi}l $  and $S = l/l_{crit}$ being a model parameter, in principle larger, or smaller
than unity.

\subsection{Numerical modelling:}

Our simulations utilize the changing Kerr metric coefficients due to the spin change and black hole growth, which we have implemented within the HARM code \citep{2003ApJ...589..444G}.We have chosen $256*256$ grid points in r and $\theta$ direction respectively. Initially fixed black hole mass is $M_{BH} = 3M_{\odot}$ .  The total mass of the surrounding gas in the cloud is $M_{cloud} = 25 M_{\odot}$ contained within the Bondi sphere of the size $R_{out} = 1000 r_{g}$ at $t=0$.

\section{Results}
In addition to metric change effects on the evolution, the mini disk formation has also been investigated and accounted for variability in
accretion rate (see \cite{2018ApJ...868...68J} for detailed results). It has been observed that matter from surrounding gas cloud accretes more rapidly into the BH with changing metric rather than the static metric during the evolution of the flow.

\begin{figure}
\includegraphics[width=\textwidth]{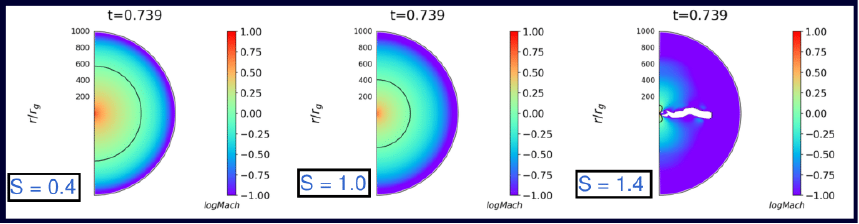}
\caption{2D Mach profile for sub critical, critical and super critical rotation has been plotted at t = 0.7 sec. Super critical rotation shows the formation of mini disk.}
\end{figure}

For super critical rotation, BH mass growth is less compared to models with sub critical and critical rotation because of the formation of mini disk during evolution of flow. This slows down feeding of BH through the inner boundary. Also due to these reasons, final BH is highly spinning with super critical rotation as compared to other rotation models. Formation of shock has been seen for critical rotation as well. Variability in accretion rate can been seen corresponding to the period of existence of shock in the flow. Further the accretion rate also drops as the shock gets accreted into BH \citep{2018ApJ...868...68J}.
\begin{figure}
\includegraphics[width=\textwidth]{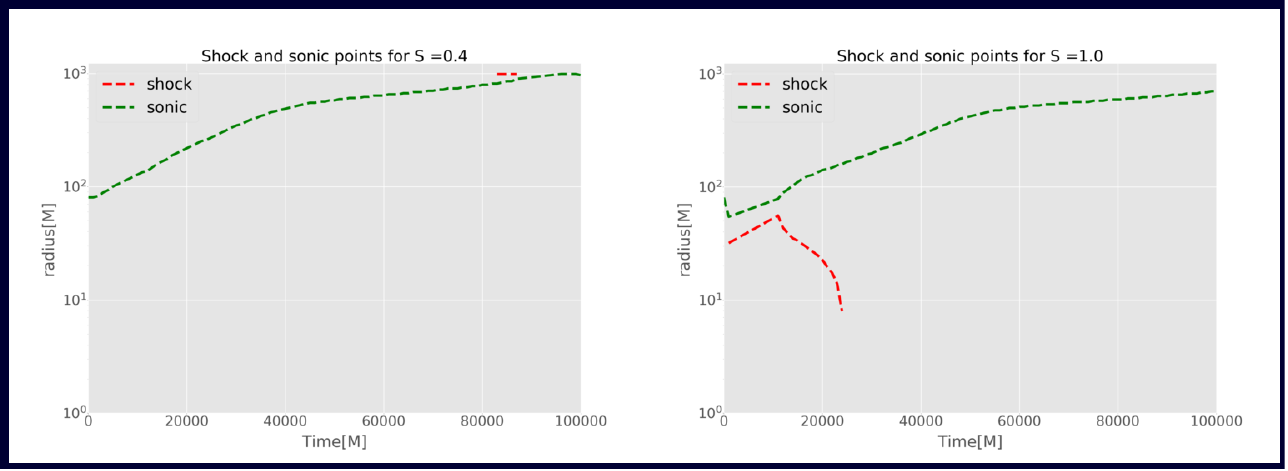}
\caption{The position of shock and sonic points with respect to time has been plotted for models with sub critical and critical rotation.The red dashed line denotes the shock formation.}
\end{figure}

\section{Summary}
We calculated GR-HD model for collapse of the star’s central parts and growth of the black hole due to accretion in changing Kerr metric.We also provided constraints for formation of a rotationally supported mini-disk in the center of collapsar.
Our results testify the fact that the massive BH detected by LIGO till date were not able to launch a powerful GRB because as per our results such massive BH should not have required high spin to support such event.

\section{Acknowledgement}
This research was supported by grant DEC-2016/23/B/ST9/03114 from the Polish National Science Center. The simulations were performed on supercomputer cluster of Interdisciplinary Center for Mathematical Modeling of the Warsaw University, under computational grant GB 79-9 and the PL-Grid computational resources through the grant grb2. PS is supported from Grant No. GACR-17-06962Y.

\bibliographystyle{ptapap}
\bibliography{palit}

\end{document}